\documentclass[conference]{IEEEtran}
\IEEEoverridecommandlockouts
\usepackage{cite}
\usepackage{amsmath,amssymb,amsfonts}
\usepackage{algorithmic}
\usepackage{graphicx}
\usepackage{textcomp}
\usepackage{listings}
\usepackage{xcolor}
\usepackage{listings}
\usepackage{booktabs}
\usepackage{diagbox}
\usepackage{url}
\usepackage{tablefootnote}
\usepackage{makecell}
\usepackage{graphicx}
\usepackage{tikz}
\usepackage{verbatim}
\usetikzlibrary{trees}
\usepackage{multirow}
\usepackage{hyperref}
\usepackage[hyphenbreaks]{breakurl}
\def\BibTeX{{\rm B\kern-.05em{\sc i\kern-.025em b}\kern-.08em
    T\kern-.1667em\lower.7ex\hbox{E}\kern-.125emX}}
\begin{document}

\title{A Framework and DataSet\\ for Bugs in Ethereum Smart Contracts
}

\author{\IEEEauthorblockN{Pengcheng Zhang$^{1}$, Feng Xiao$^{1}$, Xiapu Luo$^{2}$}
\IEEEauthorblockA{\textit{$^{1}$College of Computer and Information, Hohai University, Nanjing, P.R.China} \\
\textit{$^{2}$Department of Computing, Hong Kong Polytechnic University, HongKong, P.R.China} \\
Email: pchzhang@hhu.edu.cn; harleyxiao@foxmail.com;  csxluo@comp.polyu.edu.hk}
}

\maketitle
\begin{abstract}
Ethereum is the largest blockchain platform that supports smart contracts. Users deploy smart contracts by publishing the smart contract's bytecode to the blockchain. Since the data in the blockchain cannot be modified, even if these contracts contain bugs, it is not possible to patch deployed smart contracts with code updates. Moreover, there is currently neither a comprehensive classification framework for Ethereum smart contract bugs, nor detailed criteria for detecting bugs in smart contracts, making it difficult for developers to fully understand the negative effects of bugs and design new approaches to detect bugs. 
In this paper, to fill the gap, we first collect as many smart contract bugs as possible from multiple sources and divide these bugs into 9 categories by extending the \emph{IEEE Standard Classification for Software Anomalies}. Then, we design the criteria for detecting each kind of bugs, and construct a dataset of smart contracts covering all kinds of bugs. With our framework and dataset, developers can learn smart contract bugs and develop new tools to detect and locate bugs in smart contracts. Moreover, we evaluate the state-of-the-art tools for smart contract analysis with our dataset and obtain some interesting findings: 1) \emph{Mythril}, \emph{Slither} and \emph{Remix} are the most worthwhile combination of analysis tools. 2) There are still 10 kinds of bugs that cannot be detected by any analysis tool.  
\end{abstract}


\section{Introduction}

Millions of smart contracts have been deployed onto Ethereum, the largest blockchain that supports smart contracts. They are typically developed with  high-level programming languages and then compiled into bytecode, which will be deployed to the blockchain through transactions. Note that the deployed bytecode cannot be modified for patching the bugs. 
Unfortunately, similar to traditional computer programs, it is difficult to avoid bugs in smart contracts. Recent years have witnessed various bugs in smart contracts, resulting in huge losses. e.g., the \emph{re-entrancy bug}~\cite{atzei2017survey} in \emph{the DAO} smart contract~\cite{understandTheDAO} led to a loss of \$60 million.


Although recent studies proposed a number of tools ~\cite{luu2016making,kalra2018zeus,torres2018osiris,nikolic2018finding,chen2017under,gasreducer,gaschecker,jiang2018contractfuzzer,tikhomirov2018smartcheck,grishchenko2018ethertrust,liu2018reguard,tokenscope} for detecting the bugs in smart contracts, Ye et al.~\cite{8728953} found that they can only discover some kinds of known smart contract bugs. 
%
%
One possible reason for this situation is the lack of a comprehensive collection and classification of all existing smart contract bugs so that existing tools just aim at portions of bugs.
%
At the same time, recent studies reveal that one major reason for the prevalence of smart contract bugs is the lack of a comprehensive classification framework for smart contract bugs~\cite{atzei2017survey}. Although a few studies summarized and classified some kinds of bugs in smart contracts~\cite{atzei2017survey, 8886793,9072659,durieux2019empirical}, they have the following limitations:

\noindent$\bullet$~\emph{Existing studies do not cover all known bugs in smart contracts.} Dingman et al.~\cite{8886793} considered 49 kinds of bugs and classified them using \emph{NIST} framework. However, they grouped only 24 kinds of bugs into well-defined categories, and put the remaining 25 kinds of bugs into \emph{other} category without further classification. Moreover, they also included the bugs that have been fixed. 
Smartdec~\cite{smartdec} divided smart contract bugs into three levels: blockchain, language and model, and then classified the bugs at each level. 
Chen et al.~\cite{9072659} divided 20 kinds of bugs into 5 categories.
Durieux et al.~\cite{durieux2019empirical} divided bugs into 10 categories. However, only partial known bugs were covered by these studies.

\noindent$\bullet$~\emph{Lack of detailed bug detection criteria}. Existing work~\cite{atzei2017survey, 8886793,9072659, durieux2019empirical} only describes the causes of various bugs without giving the criteria for finding such bugs, making it difficult for developers to design algorithms and tools to detect bugs. For example, it is difficult for existing tools to detect the \emph{short address attack bug} due to the lack of detection criteria. 

\noindent$\bullet$~\emph{Existing datasets for smart contract bugs are incomplete.} e.g., \emph{SmartContractSecurity}~\cite{SWCRegistry} includes 33 kinds of bugs, but only provides sample smart contracts for 31 kinds of bugs; \emph{crytic}~\cite{NotSoSmartContracts} provides sample smart contracts that cover 12 kinds of bugs; Durieux et al.~\cite{durieux2019empirical} provide a dataset containing 69 problematic smart contracts, but they only covered 10 kinds of bugs. Without a dataset that covers all kinds of known bugs and the corresponding vulnerable smart contracts, it is difficult to comprehensively evaluate the performance of existing tools designed for finding bugs in smart contracts. 

To fill the gap, in this paper, we first carefully collect known bugs in Ethereum smart contracts from many sources, including, academic literature, the Web, blogs, and related open-source projects, and finally obtain 323 records describing bugs in Ethereum smart contract.
Then, by reviewing the \emph{Ethereum Wiki}~\cite{GithubWiki}, \emph{Ethereum Improvement Proposals}~\cite{EIP} and the development documents of \emph{Solidity}~\cite{SolidityDocument}, we remove the bugs that had been fixed by Ethereum. We also merge the bugs caused by the same behavior, and eventually get 49 kinds of bugs. After that, by extending the \emph{IEEE Standard Classification for Software Anomalies}, we classify the 49 kinds of bugs into 9 categories, and provide the detection criteria for each kind of bug. Finally, we construct a dataset of vulnerable smart contract covering all these bugs, and use it to evaluate the state-of-the-art tools for analyzing smart contracts. 
We call the framework and dataset as \emph{Jiuzhou}, which can be found at \url{https://github.com/xf97/JiuZhou}.

In summary, we make the following contributions:
\begin{itemize}
\item We propose a new framework for all known bugs in smart contracts by extending \emph{IEEE Standard Classification for Software Anomalies.} We collect these bugs from many sources and classify them into 9 categories. 
\item We design the detection criteria for each kind of bug by taking into account the cause of each bug, the most common form of bugs, and the potential false positives and negatives generated by various detection tools. 
\item We construct a dataset of problematic smart contracts, which cover all kinds of known bugs. It contains 176 smart contracts, including contracts that contain bugs, contracts with bugs fixed and crafted contracts. By studying the smart contracts in this dataset, developers and researchers can understand the patterns of these bugs and the corresponding solutions. 
\item We use our dataset as a benchmark to evaluate the state-of-the-art smart contract analysis tools. The empirical results show the detection abilities of nine smart contract analysis tools and reveal that there are still 10 kinds of bugs that cannot be detected by any of these tools. 
\end{itemize}

The rest of this paper is organized as follows: Section 2 introduces the necessary background. Section 3 presents the framework for smart contract bugs. For each bug, we not only describe its characteristics and severity level, but also design the corresponding detection criteria.
Section 4 introduces the dataset that covers all these bugs. Section 5 reports the results of using our dataset to evaluate smart contract analysis tools. After introducing the related work in Section 6, we conclude the paper with future work in Section 7. 

\section{background}

\subsection{Smart contract}
When the conditions specified in the contract are met or the smart contracts are called, smart contracts will be executed automatically on blockchain~\cite{buterin2013ethereum}. In Ethereum, each smart contract or user is assigned a unique address. Smart contracts can be invoked by sending transactions to the address of the contract\cite{chen2020understanding}. \emph{Ether} is the cryptocurrency used by Ethereum, and both contracts and users can trade \emph{ethers}. To avoid abusing the computational resources, Ethereum charges \emph{gas} from each executed smart contract statement\cite{chen2017ispec}. 

\subsection{Solidity}
Being the most widely used programming language for developing Ethereum smart contracts~\cite{SolidityDocument}, \emph{Solidity} is a Turing-complete and high-level programming language capable of expressing arbitrarily complex logic. Before deployment, the smart contracts written by \emph{Solidity} are compiled into bytecode of Ethereum virtual machine (EVM). \emph{Solidity} provides many built-in symbols to perform various functions of Ethereum. For example, \emph{transfer} and \emph{send} are used to transfer ethers, and \emph{require} and \emph{assert} are used to handle errors. \emph{Solidity} is a fast-evolving language. The same keyword may have different semantics in different versions. Fortunately, when using \emph{Solidity} to develop smart contracts, developers can specify the \emph{Solidity} version used by the contracts.

\subsection{IEEE Standard Classification for Software Anomalies} 
The \emph{IEEE Standard Classification for Software Anomalies}~\cite{isdw20101044} provides a unified method for the classification of traditional software anomalies. In the standard, \emph{error}, \emph{fault}, \emph{defect}, \emph{problem}, and \emph{bug} are uniformly described as \emph{anomalies}. In its latest version, software anomalies are classified into six categories: \emph{data}, \emph{interface}, \emph{logic}, \emph{description}, \emph{syntax}, \emph{standard}. The standard also provides ranking criteria for the effect and priority of software anomalies. It can be extended for covering different types of software. 

\section{A Classification Framework for Smart Contract Bugs}
To build a comprehensive classification framework, we collect smart contract bugs from many sources, including academic literature, the Web, blogs, and related open-source projects. Since there is no uniform bug naming standard, the same bug may have different names. Consequently, we first merge bugs according to their behaviors. Then, according to the causes of all bugs, we divided all bugs into 9 categories. Each category contains several sub-categories, and the sub-categories contain several kinds of bugs. Finally, according to the effect of different bugs, we give each bug a severity 
level.

\subsection{Collect smart contract bugs}
First, we collect smart contract bugs from academic literature, the Web, blogs, and other resources. For academic literature, we use \emph{smart contract vulnerabilities}, \emph{smart contract bugs}, \emph{smart contract defects}, \emph{smart contract problems}, and \emph{smart contract anomalies} as search keywords to search for papers published since 2014 in \emph{ACM digital library}~\cite{ACM} and \emph{IEEE Xplore digital library}~\cite{IEEE}. The reason for the paper after 2014 was chosen is that Ethereum started ICO (initial coin offering) in 2014. For the Web and blogs, we mainly focus on the \emph{Github homepage of Ethereum}~\cite{EthereumHomePage}, \emph{the development documents of Solidity}~\cite{SolidityDocument}, \emph{the official blogs of Ethereum}~\cite{EthereumBlog}, \emph{the Gitter chat room}~\cite{EthereumGitter}, \emph{Ethereum Improvement Proposals}~\cite{EIP} and other resources. 

Second, the open-source projects are also our focus since the open-source community plays an important role in the field of software security~\cite{parizi2018empirical}. Specifically, we use \emph{smart contract bugs}, \emph{smart contract problems}, \emph{smart contract defects}, \emph{smart contract vulnerabilities}, and \emph{smart contract anomalies} as search keywords to retrieve related open-source projects on \emph{GitHub}~\cite{Github}. 
Besides, many smart contract analysis tool projects are also open-sourced on \emph{GitHub}~\cite{Github}, and some of the project documents describe information about smart contract bugs. Therefore, we also use \emph{smart contract analysis tools} and \emph{smart contract security} as search keywords to retrieve open-source projects on \emph{Github}. We focus on the projects for Ethereum smart contracts. After removing duplicate search results, we obtain a total of 266 projects. 

Third, since some famous Ethereum smart contract analysis tools can detect smart contract bugs, we sent emails to the authors of these tools asking what kinds of bugs they can detect. We also look at the kinds of bugs detected by the \emph{Solidity static analysis} function of \emph{Remix}~\cite{Remix}. Finally, from the resources mentioned above, we collected 323 records describing Ethereum smart contract bugs.

To continuously collect bugs,  
we develop a program called \emph{BugGetter\footnote{https://github.com/xf97/BugGetter}}, which runs regularly (by default 15 days) and sends query requests to \emph{Github}. \emph{BugGetter} uses keywords such as \emph{smart contract vulnerabilities}, \emph{smart contract bugs}, \emph{smart contract defects}, \emph{smart contract problems}, \emph{smart contract security}, and \emph{smart contract analysis tools} to construct query requests to \emph{Github}, and 
extract the list of projects and their update time from the response. 
By comparing previously obtained projects and their update time, \emph{BugGetter} will send us an email if a new project appears or an existing project is updated. After receiving the email, we will manually check all changes and update the collected bug results in time.

\subsection{Merge smart contract bugs}
Since there is no uniform bug naming standard, some bugs are actually the same even if they have different names. 
Therefore, we need to merge the duplicate bugs. The collected bugs generally have two attributes, namely, the behaviors causing the bug and the consequences caused by the bug. If there is a bug \emph{A}. 
Let,
\begin{itemize}
\item the behaviors causing bug \emph{A} be \emph{b(A)},
\item the consequences caused by bug \emph{A} be \emph{c(A)}.
\end{itemize}

Given two bugs \emph{A} and \emph{B}, we handle them as follows: 1) If $ b(A) \ne b(B) $, we do not merge \emph{A} and \emph{B}. 2) If $ b(A) = b(B), c(A) \ne c(B)$, \emph{c(A)} and \emph{c(B)} cover part of the consequences of the bug. Hence, we merge \emph{A} and \emph{B}, rename the merged bug, summarize \emph{c(A)} and \emph{c(B)} as the consequences caused by the merged bug. 3) If $ b(A) = b(B), c(A) = c(B)$, we merge \emph{A} and \emph{B} and choose the name that better characterizes the bug as the name of the merged bug. 

After the duplicate bugs are merged, we verify the validity of each bug (i.e., the bug has not been permanently fixed by Ethereum), and delete the fixed bugs. Finally, 49 kinds of bugs are left. The bugs before and after the merging step can be found at \emph{\url{https://github.com/xf97/JiuZhou/blob/master/Correspondence.xlsx}}. 

\subsection{Classify smart contract bugs}

\noindent\emph{\textbf{C.1 Classification criteria and results:}}

According to \emph{IEEE Standard Classification for Software Anomalies~\cite{isdw20101044}} issued in 2010, software anomalies are classified into six categories: \emph{data}, \emph{interface}, \emph{logic}, \emph{description}, \emph{syntax}, \emph{standard}. Among them, we do not consider the \emph{syntax} category, because a smart contract with syntax bugs can be neither compiled into bytecode nor deployed to Ethereum.
Note that some bugs are specific to Ethereum, such as  the bugs caused by \emph{gas}, \emph{lack of privacy on the blockchain}, \emph{smart contract authority control}, \emph{smart contract interactions}, and \emph{smart contract support software}. 
Since they cannot be accurately classified by \emph{IEEE Standard Classification for Software Anomalies~\cite{isdw20101044}}, 
%
we add four new categories: \emph{security (for lack of privacy and authority control)}, \emph{performance (for gas consumption)}, \emph{interaction (for smart contract interaction and ethers exchange)}, and \emph{environment (for smart contract support software)}. Consequently, we divide smart contract bugs into the following nine categories lexicographically:

\begin{enumerate}
    \item \textbf{\emph{Data}}. Bugs in data definition, initialization, mapping, access, or use, as found in a model, specification, or implementation.
    \item \textbf{\emph{Description}}. Bugs in the description of the software or its use, installation, or operation.
     \item \textbf{\emph{Environment}}. Bugs due to errors in the supporting software.
     \item \textbf{\emph{Interaction}}. Bugs that cause by interaction with other Ethereum addresses.
    \item \textbf{\emph{Interface}}. Bugs in specification or implementation of an interface.
    \item \textbf{\emph{Logic}}. Bugs in decision logic, branching, sequencing, or a computational algorithm, as found in natural language specifications or implementation language. 
    \item \textbf{\emph{Performance}}. Bugs that cause increased \emph{gas} consumption.
    \item \textbf{\emph{Security}}. Bugs that threaten contract security, such as authentication, privacy/confidentiality, property.
    \item \textbf{\emph{Standard}}. Nonconformity with a defined standard.
\end{enumerate}

When merging bugs, we check \emph{Ethereum Improvement Proposals}~\cite{EIP}, \emph{the Ethereum Wiki}~\cite{GithubWiki}, and the development documents of \emph{Solidity}~\cite{SolidityDocument} to remove the bugs that have been fixed by Ethereum (e.g., the \emph{call depth attack}, which was fixed in \emph{EIP150}~\cite{EIP150}). 
Some kinds of bugs are caused by specific \emph{Solidity} versions. Since it is still possible to use these versions of \emph{Solidity} to develop smart contracts, we list the range of \emph{Solidity} versions that cause these kinds of bugs. The remaining kinds of bugs exist in any version of \emph{Solidity}.

\begin{figure*}[htbp]
    \centering
    \includegraphics[scale=0.31]{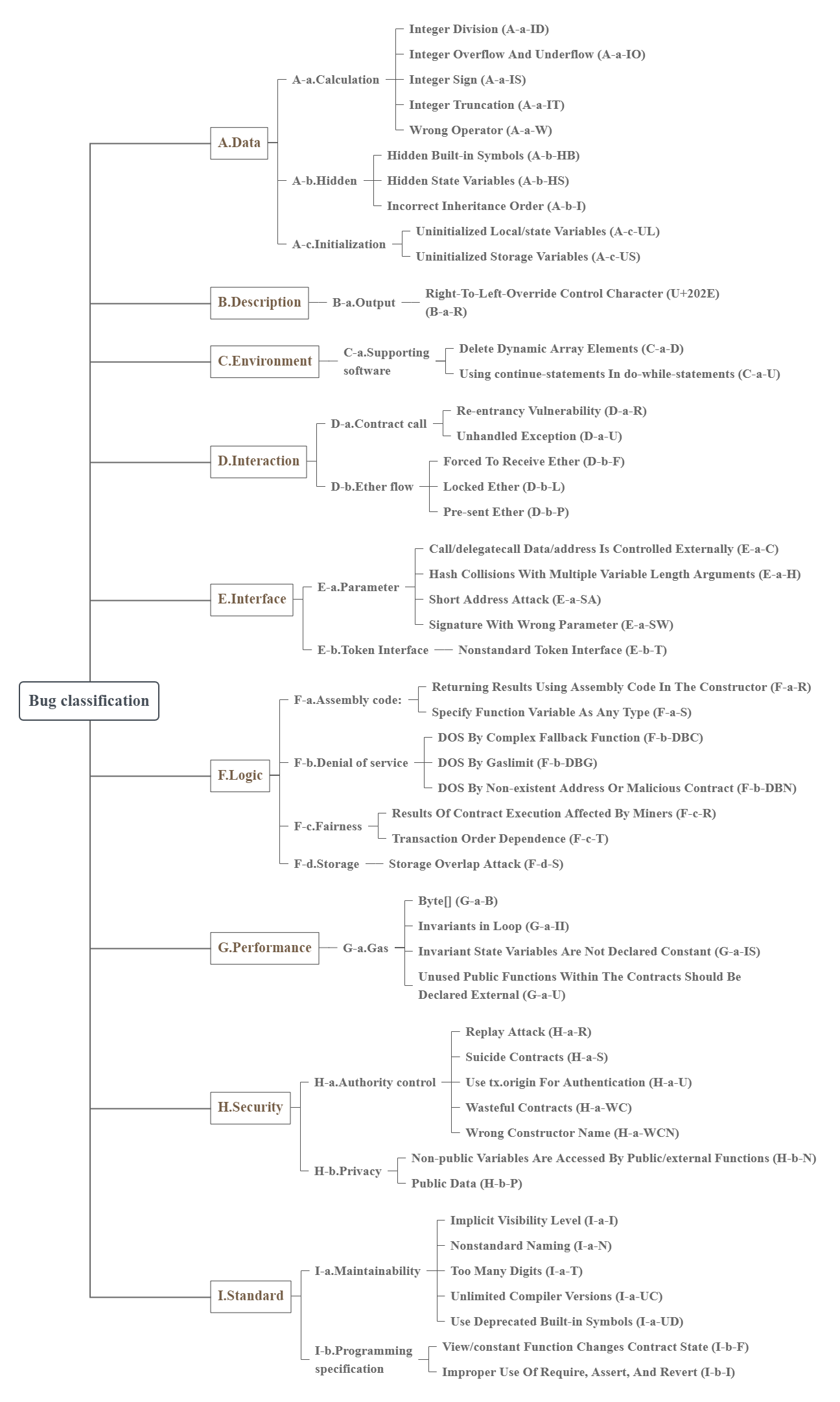}
    \caption{Smart contract bug statistics and classification}
    \label{fig:classification}
\end{figure*}

We divide 49 kinds of bugs into 9 categories, each of which is divided into several sub-categories, and each sub-category contains several kinds of bugs. Fig~\ref{fig:classification} shows the classification of 49 kinds of bugs. For ease of expression, we assign a corresponding abbreviated name to each bug. The composition rule of the abbreviated name is: \emph{category number-subcategory number-short name}. The \emph{short name} is an acronym that can be distinguished from the \emph{short names} of other kinds of bugs in the same sub-category. Since classifying bugs based on their descriptions is a manual process based on natural language descriptions, it may introduce the subjectivity and ambiguity. To mitigate the impact of this problem, three researchers who are familiar with smart contract bugs participated in the classification process and reached a consensus through discussion.
Due to space limit, in this paper, we only introduce some kinds of bugs in detail. For the full version, please visit this \emph{url\footnote{\url{https://github.com/xf97/JiuZhou/blob/master/Jiuzhou_Full_version.pdf}}}.\newline

\noindent\emph{\textbf{C.2 Selected kinds of smart contract bugs}}

In this part, we introduce the causes, consequences, and detection criteria of seven kinds of bugs in detail, and illustrate these bugs with examples. Some of these bugs are difficult to understand the attack process while others are not mentioned in other studies~\cite{atzei2017survey, 8886793,9072659}. Table~\ref{tab:bug_detail} lists these bugs.

\begin{table}[h]
    \centering
    \caption{7 kinds of bugs introduced in detail}
    \label{tab:bug_detail}
    \begin{tabular}{c|c}
    \toprule
    \textbf{\makecell{Bug name \\abbreviation}} & \makecell{A-a-IS, A-a-W, A-c-US, D-a-R\\ E-a-SA, E-a-SW, F-c-T}\\
    \bottomrule
    \end{tabular}
\end{table}

\noindent\textbf{Integer Sign} (A-a-IS)~\cite{torres2018osiris}:
\begin{itemize}
    \item \emph{Cause}:  In \emph{Solidity}, Converting \emph{int} type to  \emph{uint} type (and vice versa) may produce incorrect results.
    \item \emph{Consequence}: This kind of bugs may result in incorrect integer operation results, which will affect the function of the contract. When the wrong result is used to indicate the number of ethers (or tokens), this kind of bugs will cause economic losses.
    \item \emph{Example}: Consider the smart contract in Fig~\ref{fig:eg_IS}, if the attacker calls the \emph{withdrawOnce} function and specifies that the value of \emph{amount} is a negative number, then this call will pass the check and transfer out the ether that exceeds the limit (1 ether).
    \begin{figure}[ht]
        \centering
        \includegraphics[scale=0.6]{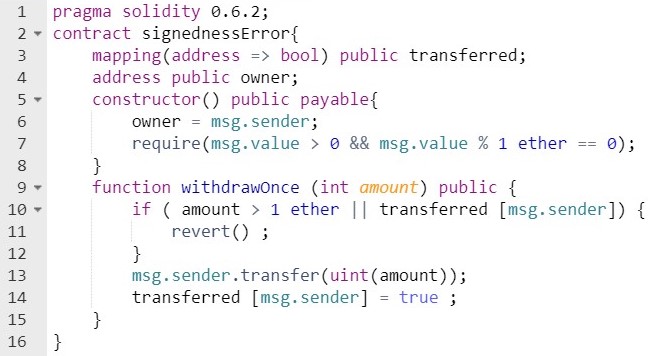}
        \caption{A contract that contains \emph{integer sign} bug}
        \label{fig:eg_IS}
    \end{figure}
    \item \emph{Detection criteria}: This kind of bugs exists when the  contract contains the following features: 1) Forcibly convert an \emph{int} variable  to a \emph{uint} variable. 2) The contract does not check whether this \emph{int} variable is negative.
\end{itemize}

\noindent\textbf{Wrong Operator} (A-a-W)~\cite{SWCRegistry}:
\begin{itemize}
    \item \emph{Cause}: Users can use =+ and =- operators in the integer operation without compiling errors (up to and including version 0.4.26).
    \item \emph{Consequence}: Consistent with the consequences of \emph{integer sign} bug.
    \item \emph{Example}: Consider the smart contract in Fig~\ref{fig:eg_WO}, the user can adjust the value of \emph{myNum} by calling \emph{addOne}/\emph{subOne} function. When \emph{myNum} and \emph{WinNum} are equal, the user gets all the ethers of the contract. But because the developers write the wrong operators (Line 15, 19), the value of \emph{myNum} will not be changed.
    \begin{figure}
        \centering
        \includegraphics[scale=0.6]{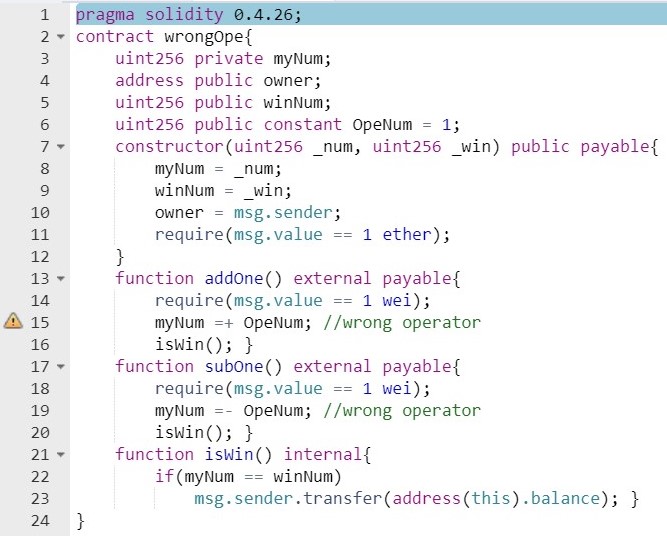}
        \caption{A contract contains a \emph{wrong operator} bug}
        \label{fig:eg_WO}
    \end{figure}
    \item \emph{Detection criteria}: This kind of bugs exists when the following feature exists in the contract: 1) There is a =+ or =- operator in the contract.
\end{itemize}

\noindent\textbf{Uninitialized Storage Variables} (A-c-US)~\cite{SWCRegistry}: 
\begin{itemize}
    \item \emph{Cause}: The uninitialized storage variable serves as a reference to the first state variable in a contract, which may cause the state variable to be inadvertently modified (up to and including version 0.4.26).
    \item \emph{Consequence}: This kind of bugs may cause key state variables to be rewritten inadvertently, and eventually, the function of the contract will be affected.
    \item \emph{Example}: Consider the smart contract in Fig~\ref{fig:eg_US}, when the user calls the function \emph{func}, the \emph{owner} will be re-assigned to 0x0.
    \begin{figure}[ht]
        \centering
        \includegraphics[scale=0.6]{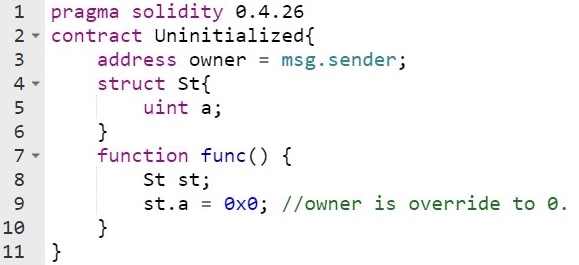}
        \caption{A contract contains a \emph{uninitialized storage variables} bug}
        \label{fig:eg_US}
    \end{figure}
    \item \emph{Detection criteria}: This kind of bugs exists when the contract contains the following features: 1) the developers do not initialize the storage variables in the contract.
\end{itemize}

\noindent\textbf{Re-entrancy Vulnerability} (D-a-R)~\cite{siegel2016understanding, liu2018reguard}: 
\begin{itemize}
    \item \emph{Cause}: When the \emph{call}-statement is used to call other contracts, the callee can call back the caller and enter the caller again.
    \item \emph{Consequence}: This kind of bugs is one of the most dangerous smart contract bugs, which will cause the contract balance (ethers) to be stolen by attackers.
    \item \emph{Example}: We use an example to illustrate the \emph{re-entrancy vulnerability}. In Fig~\ref{fig:re}, the contract \emph{Re} is a contract with a \emph{re-entrancy vulnerability}, and the \emph{balance} variable is a map used to record the correspondence between the address and the number of tokens. The attacker deploys the contract \emph{Attack}, and the value of the parameter \emph{\_reAddr} is set to the address of the contract \emph{Re}.
	\begin{figure*}[htp]
\centering
\includegraphics[scale=0.33]{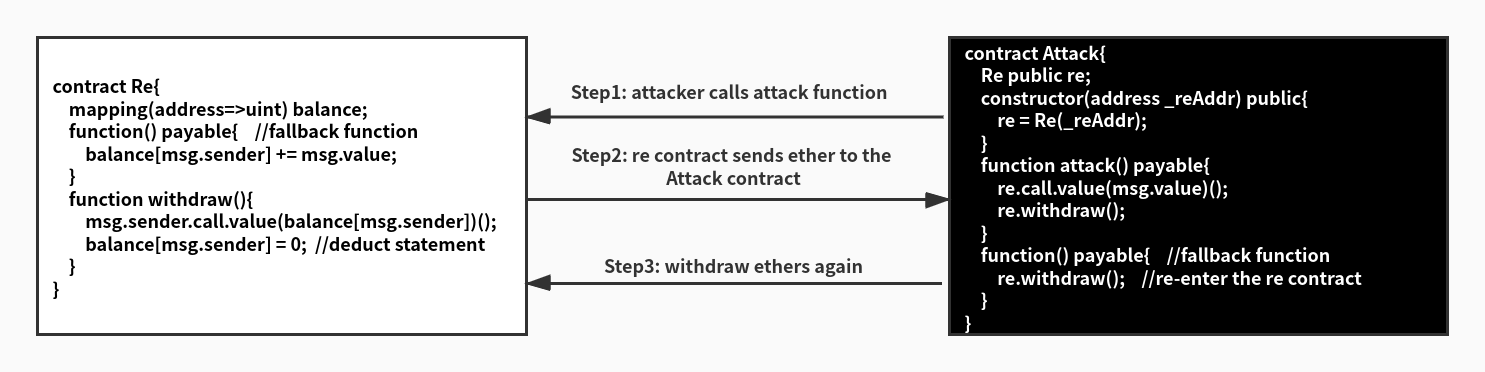}
\caption{An example of \emph{re-entrancy vulnerability}}
\label{fig:re}
\end{figure*}
	In this way, the \emph{re} variable becomes an instance of the contract \emph{Re}. Then,
		\begin{itemize}
		\item \emph{Step 1}: The attacker calls the \emph{attack} function to deposit ethers into the contract \emph{Re} and then calls the \emph{Re.withdraw} function to retrieve the deposited ethers.
		\item \emph{Step 2}: The contract \emph{Re} executes the \emph{withdraw} function and uses a \emph{call}-statement to send ethers to the contract \emph{Attack}. At this time, the power of control is transferred to the contract \emph{Attack}, and the contract \emph{Attack} responds to the transfer using the \emph{Attack.fallback} function.
		\item \emph{Step 3}: The \emph{Attack.fallback} function calls the \emph{Re.withdraw} function to withdraw the ethers again. Therefore, the statement (\emph{deduct-statement}) deducting the number of tokens held by the contract \emph{Attack} will not be executed.
		\end{itemize}
    \item \emph{Detection criteria}: When the following features are in the contract, it will cause the \emph{reentrancy bug}: 1) The \emph{call-statement} is used to send ethers. 2) The amount of \emph{gas} to be carried is not specified. 3) No callee's response function is specified. 4) Ethers are transferred first and callee's balance is deduced later. 

\end{itemize}

\noindent\textbf{Short Address Attack} (E-a-SA)~\cite{doingblock}: 
\begin{itemize}
    \item \emph{Cause}: When Ethereum packs transaction data if the data contains the address type and the length of the address type is less than 20 bits, subsequent data will be used to make up the length of the address type.
    \item \emph{Consequence}: An attacker can exploit this kind of bug to manipulate much more tokens (Ethers) than he requested.
    \item \emph{Example}: We use an example to illustrate \emph{short address attacks}.
        \begin{itemize}
            \item \emph{Step 1}: \emph{Tom} deploys token contract \emph{A} on Ethereum, which contains the \emph{sendCoin} function. The code of \emph{sendCoin} is shown in Fig~\ref{fig:shortAttack}.
            \begin{figure}[htbp]
            \centering
            \includegraphics[scale=0.47]{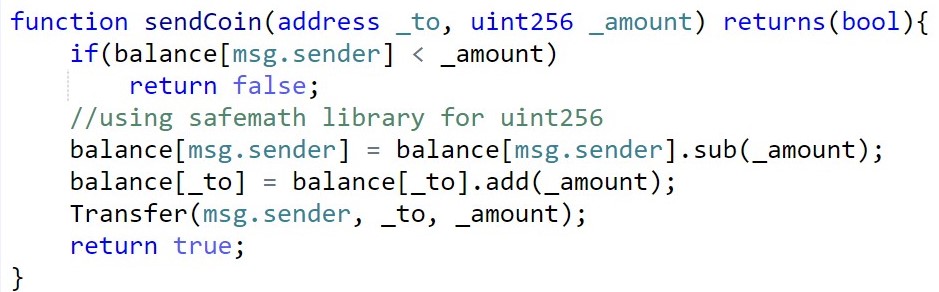}
            \caption{Objective function of \emph{short address attack}}
        \label{fig:shortAttack}
    \end{figure}
        \item \emph{Step 2}: \emph{Jack} buys 100 tokens of contract \emph{A}, then registers for an Ethereum account with the last two digits zero (e.g. 0x1234567890123456789012345678901234567800).
        \item \emph{Step 3}: \emph{Jack} calls the function \emph{sendCoin} with the given parameters, \emph{\_to}: 0x12345678901234567890123456789012345678 (missing last two digits 0), \emph{\_amount}: 50.
        \item \emph{Step 4}: The value of \emph{\_amount} is less than 100, so it passes the check. However, because the bits of \emph{\_to} is insufficient, the first two bits (0) of \emph{\_amount} will be added to the \emph{\_to} when the transaction data is packed. Therefore, to make up for the length of \emph{\_amount}, the Ethereum virtual machine will add 0 to the last two bits. In the end, the value of \emph{\_amount} is expanded by four times.
        \end{itemize}
    \item \emph{Detection criteria}: When there are the following features in the contract, it will cause the \emph{short address attack}: 1) The contract uses a function to transfer ethers or tokens. 2) The number of tokens (ethers) and the address for receiving tokens (ethers) are provided by external users. 3) There is no operation to check the length of the received tokens (ethers) address in the function.

\end{itemize}

\begin{figure}[htbp] 
\centering
\includegraphics[scale=0.55]{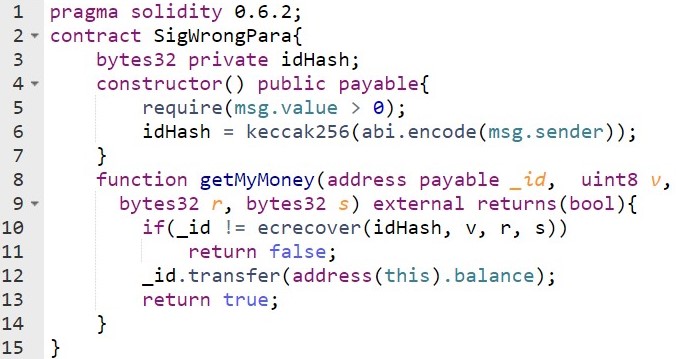}
\caption{A contract contains a \emph{signature with wrong parameter} bug}
\label{fig:eg_WP}
\end{figure}

\noindent\textbf{Signature With Wrong Parameter} (E-a-SW)~\cite{secBit}:  
\begin{itemize}
    \item \emph{Cause}: When the parameters of the \emph{ecrecover()} are wrong, the \emph{ecrecover()} will return 0x0.
    \item \emph{Consequence}: This kind of bugs will allow the attacker to pass the authentication and then the attacker can manipulate the token (ethers) held by the \emph{0x0} address.
    \item \emph{Example}: Considering the smart contract in Fig~\ref{fig:eg_WP}, when the attacker gives the wrong parameters (\emph{v}, \emph{r}, \emph{s}) and the value of the specified parameter \emph{\_id} is \emph{0x0}, the attacker can pass the identity verification (Line 10), which eventually leads to the ethers in the contract are destroyed.
    \item \emph{Detection criteria}: This kind of bugs exists when the contract contains the following features: 1) There is an operation in the contract that uses the \emph{ecrecover()} to calculate the public key address. 2) The contract does not deal with the case where the \emph{ecrecover()} returns \emph{0x0}.
\end{itemize}

\noindent\textbf{Transaction Order Dependence} (F-c-T)~\cite{tsankov2018securify}:
\begin{itemize}
    \item \emph{Cause}: Miners can decide which transactions are packaged into the blocks and the order in which transactions are packaged. The current main impact of this kind of bugs is the \emph{approve} function in the \emph{ERC20} token standard.
    \item \emph{Consequence}: This kind of bugs will enable miners to influence the results of transaction execution. If the results of the previous transactions will have an impact on the results of the subsequent transactions, miners can influence the results of transactions by controlling the order in which the  transactions are packaged.
    \item \emph{Example}: The \emph{approve} function allows one address to approve another address to spend tokens on his behalf. The standard implementation of the \emph{approve} function is shown in Fig~\ref{fig:eg_TO}. We assume that \emph{Alice} and \emph{Tom} are two Ethereum users, and \emph{Tom} runs an Ethereum node. The following steps reveal how \emph{Tom} uses the \emph{transaction order dependence} bug to monetize:
    \begin{itemize}
        \item \emph{Step 1}: Assume \emph{Alice} has approved \emph{Tom} to spend $n$ of the tokens she holds. Now \emph{Alice} decides to change \emph{Tom}'s quota to $m$ tokens, so \emph{Alice} sends a transaction to modify \emph{Tom}'s quota.
        \item \emph{Step 2}: Since \emph{Tom} runs an Ethereum node, he knows that \emph{Alice} will change his quota to $m$ tokens. Then, \emph{Tom} sends a transaction (e.g., Using the \emph{transferFrom} function of the \emph{ERC20} token standard to transfer $n$ tokens to himself) to spend \emph{Alice}'s $n$ tokens, and pays a lot of \emph{gas} to make his transaction executed first.
        \item \emph{Step 3}: The node that obtains the accounting right packs transactions. Because \emph{Tom} pays more \emph{gas}, \emph{Tom}'s transaction will be executed before \emph{Alice}'s transaction. Therefore, \emph{Tom} spent $n$ tokens of \emph{Alice} first and then is granted a quota of $m$ tokens by \emph{Alice}, which caused \emph{Alice} to suffer losses.
    \end{itemize}
    \begin{figure}
        \centering
        \includegraphics[scale=0.6]{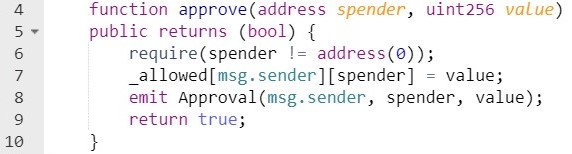}
        \caption{The standard implementation of the \emph{approve} function in the \emph{ERC20} token standard.}
        \label{fig:eg_TO}
    \end{figure}
    \item \emph{Detection criteria}: This kind of bugs exists when the contract contains the following features: 1) The contract contains the \emph{approve} function of the \emph{ERC20} token standard. 2) In the \emph{approve} function, the quota of the approved address is set from one nonzero value to another nonzero value.
\end{itemize}

\subsection{Severity level of smart contract bugs}
To help developers and researchers have a clear understanding of the consequence of each bug, we grade the severity of each kind of bug. According to the \emph{IEEE Standard Classification for Software Anomalies\cite{isdw20101044}}, we classify the effect of bugs into the following four categories:
\begin{itemize}
    \item \emph{\textbf{Functionality}}. The required function cannot be performed correctly (or an unwanted function is performed).
    \item \emph{\textbf{Performance}}. Failure to meet performance requirements, such as rising operating costs.
    \item \emph{\textbf{Security}}. Failure to meet security requirements, such as failure of authority control, privacy breaches, property theft, etc.
    \item \emph{\textbf{Serviceability}}. Failure to meet maintainability requirements, such as reduced code readability.
\end{itemize}

According to the harmfulness of the above four effects, the grading criteria of these bugs are described as follows:
\begin{itemize}
	\item \textbf{Critical}: These kinds of bugs must affect security.
	\item \textbf{High}: These kinds of bugs may affect security or necessarily affect functionality.
	\item \textbf{Middle}: These kinds of bugs may affect the functionality or necessarily affect performance.
	\item \textbf{Low}: These kinds of bugs may affect performance or necessarily affect serviceability.
\end{itemize}

According to the grading criteria, the severity level of each kind of bug is shown in Table~\ref{tab:severity}.

\begin{table}
\centering
  \caption{A classification of severity levels of each kind of bug}
  \label{tab:severity}
\begin{tabular}{cc|cc|cc}
    \toprule
    \textbf{Name} & \textbf{Severity} & \textbf{Name} & \textbf{Severity} & \textbf{Name} & \textbf{Severity} \\
	\midrule
	A-a-ID & High & 	A-a-IO & High & A-a-IS & High  \\
	\midrule 
	A-a-IT & High & 	A-a-W & High  & A-b-HB & High \\
	\midrule
	A-b-HS & High & 	 A-b-I & High & A-c-UL & High \\
	\midrule
	A-c-US & High & 		B-a-R & Middle & C-a-D & Middle \\
	\midrule
	C-a-U & Middle & 	D-a-R & Critical & D-a-U & High \\
	\midrule 
	D-b-F & Middle & 		D-b-L & Critical  & D-b-P & Middle \\
	\midrule
	E-a-C & High &	E-a-H & High  & E-a-SA & High \\
	\midrule
		E-a-SW & High & 	E-b-T & High & F-a-R  & Middle \\
	\midrule
	F-a-S & Middle &	\makecell{F-b\\-DBC} & Middle & \makecell{F-b\\-DBG} & Middle \\
	\midrule
	\makecell{F-b\\-DBN} & High & 	F-c-R & Middle  & F-c-T & Middle \\
	\midrule
	F-d-S & High & 	G-a-B  & Low & 	G-a-II & Middle \\
	\midrule 
	 G-a-IS &  Middle  &	G-a-U & Middle & 	H-a-R & High \\
	\midrule
	 H-a-S & Critical &	 H-a-U & High & 	H-a-WC & Critical \\
	\midrule
	\makecell{H-a-\\WCN} & High &H-b-N & High & H-b-P & High   \\
	\midrule
I-a-I & Low &	I-a-N & Low  & I-a-T  & High \\
	\midrule
	I-a-UC & Low &	I-a-UD & Low  & I-b-I & Middle \\
	\midrule
    I-b-F & Middle & 	 &  & 	 &  \\
  \bottomrule
\end{tabular}
\end{table}

\section{\emph{Jiuzhou}: a dataset of smart contracts with bugs}

\subsection{Overview}
\emph{Jiuzhou} provides examples of each kind of bugs to help smart contract researchers and developers better understand the bugs. The dataset can be used as a benchmark to evaluate the abilities of smart contract analysis tools. 
\emph{Jiuzhou} includes 176 smart contracts, covering all smart contract bugs studied in this paper. For each kind of bug, \emph{Jiuzhou} provides at least a contract with the bug and a contract without the bug. We also manually prepare some contracts that may cause existing analytical tools to make mistakes. 
%
The reason for providing such contracts is that the test cases used in the existing studies to evaluate the abilities of smart contract analysis tools usually only contain bugs without \emph{"bug-like"} statements that may lead to false positives. 
Therefore, those tools have not been thoroughly evaluated~\cite{parizi2018empirical}~. 
%
To address this issue, we use the following strategies to construct the crafted contracts that may mislead the smart contract analysis tools, and include them in our dataset.

\begin{itemize}
	\item \textbf{Premise}: According to whether detecting a bug relies on its context, we divide the bugs into context-independent bugs and context-dependent bugs. The former can be located according to the specific statements (e.g., \emph{nonstandard naming}, \emph{byte[]}, \emph{wrong operator}) whereas detecting the latter needs to take into account the context statements of the bug. We only create crated contracts for context-dependent bugs.
	%
	\item \textbf{Strategy 1}: This strategy divides the statement (or structure) that causes the bug into multiple statements or multiple functions or multiple contracts. It can effectively reduce the precision of static code scanning tools. 
	\item \textbf{Strategy 2}: This strategy provides the statements (or structures) that can fix the bug, but make the statements (or structures) invalid or unreachable. It can induce analysis tools to miss the bug.
	\item \textbf{Strategy 3}: This strategy uses uncommon means to fix the bug. It can mislead analysis tools to report a false positive.
\end{itemize}

\subsection{Smart contract sources}
We collect smart contracts from the following three sources:
\begin{itemize}
    \item Other smart contract datasets (e.g.,~\cite{SWCRegistry},  ~\cite{NotSoSmartContracts}). Since most of these contracts are developed using some old version of \emph{Solidity}, we manually rewrite these smart contracts using the latest version of \emph{Solidity} and keep the bugs.
    \item Sample code in the papers (e.g.~\cite{torres2018osiris}), or sample code for smart contract audit checklists (e.g.~\cite{404Checklist}). Since most of the sample code only contains one function or part of the contract, we construct a complete smart contract to include these sample codes.
    \item We manually develop smart contracts based on the features of some kinds of bugs. For the bugs that have only text descriptions without sample code, we prepare smart contracts that contain them manually and assure that the bugs can be trigger and cause the expected consequences.
\end{itemize}

Table~\ref{tab:source_num} shows the distribution of the number of unchanged smart contracts, modified smart contracts, and smart contracts that are developed by ourselves.
 
 \begin{table}[h]
 \centering
  \caption{Numbers of three kinds of smart contracts}
  \label{tab:source_num}
  \begin{tabular}{c|c|c|c}
    \toprule
    ~ & \makecell{\textbf{Unchanged} \\ \textbf{smart contracts}} & \makecell{\textbf{Modified}\\ \textbf{smart contracts}} & \makecell{\textbf{Handwritten}\\ \textbf{smart contract}} \\
	\midrule
	Num & 21 & 69 & 86 \\
  \bottomrule
\end{tabular}
\end{table}

\subsection{Comparison with other datasets}
All problematic smart contracts in the \emph{Jiuzhou} dataset were developed using the latest version (0.4.26, 0.5.16, or 0.6.2) of \emph{Solidity}. Compared with several commonly used smart contract datasets~\cite{ethernaut,NotSoSmartContracts,SWCRegistry, capture}, \emph{Jiuzhou} provides more smart contracts, uses the latest versions of \emph{Solidity}, and covers more kinds of smart contract bugs. A comparison of \emph{Jiuzhou} with other commonly used datasets is shown in Table~\ref{tab:dataset_compare}.

\begin{table}
\centering
  \caption{Comparison of \emph{Jiuzhou} with other datasets}
  \label{tab:dataset_compare}
  \begin{tabular}{c|c|c|c}
    \toprule
    \textbf{\makecell{dataset}} & \textbf{\makecell{Number of \\ contracts}} & \textbf{\makecell{Kinds of \\ bugs}}  & \textbf{\emph{Solidity version}}\\
	\midrule
	\emph{Jiuzhou} & 176 & 49 & 0.4.26 to 0.6.2 \\
	\midrule
	\emph{ethernaut}~\cite{ethernaut} & 21 & 21 & 0.4.18 to 0.4.24 \\
	\midrule
	\emph{\makecell{not-so-smart\\-contracts}}~\cite{NotSoSmartContracts} & 25 & 12 & 0.4.9 to 0.4.23 \\
	\midrule
	\emph{SWC-registry}~\cite{SWCRegistry}& 114 & 33 & 0.4.0 to 0.5.0 \\
	\midrule
	\emph{capturetheether}~\cite{capture} & 19 & 6 & 0.4.21 \\ 
  \bottomrule
\end{tabular}
\end{table}

\subsection{Usages of the \emph{Jiuzhou} dataset}
Our dataset are useful to different kinds of developers:
\begin{itemize}
    \item For developers of smart contracts, they can learn the smart contract bugs by reading these smart contracts. 
    \item For developers of smart contract analysis tools, the \emph{Jiuzhou} dataset can guide them to develop smart contract analysis tools. They can learn the patterns of problematic smart contracts that are prone to false positives or false negatives. 
    \item For users who want to evaluate smart contract analysis tools, they can use these smart contracts as a benchmark to evaluate the abilities of smart contract analysis tools. 
\end{itemize}

\section{Evaluation of smart contract analysis tools}
\subsection{Overview}
We use smart contracts in the \emph{Jiuzhou} dataset as a benchmark to evaluate the abilities of several smart contract analysis tools. An analysis tool with good ability should be able to analyze as many kinds of bugs as possible, and it should also have good precision and recall rate. Therefore, we use the following indicators to measure the abilities of the analysis tools:
\begin{itemize}
    \item \textbf{Coverage}. Coverage refers to the proportion of various bugs that can be detected by the analysis tool in the various bugs of \emph{Jiuzhou} statistics. e.g., \emph{Oyente} claims to be able to detect 3 kinds of bugs (\emph{A-a-IO}, \emph{D-a-R}, \emph{F-c-T}), and these three kinds of bugs are in \emph{Jiuzhou} statistics, so \emph{Oyente}'s coverage rate is 6\% (3/49).
    \item \textbf{Precision and recall}. We use equation~\ref{equ:precision} and equation~\ref{equ:recall} to calculate the precision and recall rate. \emph{tp} means that the tool analyzes the existence of the bug and the bug does exist. \emph{fp} means that the tool analyzes the existence of the bug but the bug does not exist. \emph{fn} means that the bug exists but the tool does not report the bug. The definitions of \emph{tp}, \emph{fp}, and \emph{fn} are shown in Table~\ref{tab:score_define}.
    \begin{table}[h]
\centering
  \caption{Definition of \emph{tp}, \emph{fp}, \emph{fn}}
  \label{tab:score_define}
  \begin{tabular}{c|c|c}
    \toprule
     \diagbox{Actual}{Analysis} & \textbf{exist} & \textbf{non-exist}  \\
     \midrule
      \textbf{exist} & \emph{tp} & \emph{fn} \\
      \midrule
      \textbf{non-exist} & \emph{fp} & ~ \\ 
  \bottomrule
\end{tabular}
\end{table}
\end{itemize}

\begin{equation}\label{equ:precision}
Precision =  (tp\div(tp+fp))
\end{equation}

\begin{equation}\label{equ:recall}
Recall =  (tp\div(tp+fn))
\end{equation}

We select and evaluate the tools according to the following two criteria: 1) The tool is free to use. 2) The tool takes a \emph{Solidity} contract or compiled bytecode as input.

Table~\ref{tab:tools} lists the 9 smart contract analysis tools we selected. To the best of our knowledge, the number of tools evaluated in this paper is not less than any existing work~\cite{durieux2019empirical, parizi2018empirical, asem2020how}. When installing these tools, we use the quick (or easy) method provided by the tools to install.

\begin{table}[htp]
    \centering
        \caption{The selected nine smart contract analysis tools for evaluation}
    \label{tab:tools}
    \begin{tabular}{c|c}
    \toprule
        \textbf{Tool}  & \makecell{Maian~\cite{MaianTool}, Mythril~\cite{Mythril}, Osiris~\cite{OsirisTool}, Oyente~\cite{OyenteTool}, Securify~\cite{SecurifyTool}, \\ Slither~\cite{SlitherTool},  SmartCheck~\cite{SmartCheckTool}, Remix\tablefootnote{We use the \emph{solidity static analysis} of Remix}~\cite{RemixIDE}, SolidityCheck~\cite{SolidityCheckTool}}\\
        \bottomrule
    \end{tabular}
\end{table}

\subsection{Coverage}
We obtain the kinds of bugs that can be detected by a tool according to its documents. 
For the tools without detailed documents, we ask the developers via email. 
Fig~\ref{fig:exp_data} shows the coverage of various tools. The coverage of \emph{Slither} is the highest, and the coverage of static code scanning tools (e.g., \emph{SmartCheck}, \emph{SolidityCheck}) is usually high. But the coverage of tools based on control flow (or data flow) analysis is usually low.

\subsection{Precision and recall}
\begin{figure}[ht]
    \centering
    \includegraphics[scale=0.61]{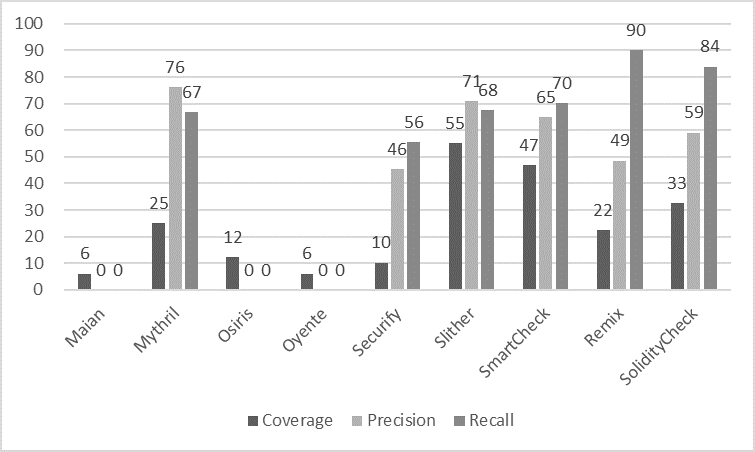}
    \caption{\emph{Coverage}, \emph{recall} and \emph{precision} of various tools}
    \label{fig:exp_data}
\end{figure}

The test cases we chose are the smart contracts containing all kinds of bugs that a tool claims to be able to detect. 
We instruct each analysis tool to analyze these contracts, and then calculate the recall and precision of each tool. 
The results are shown in Fig~\ref{fig:exp_data}. Nine of the evaluated tools including six bytecode-based tools (\cite{MaianTool}, \cite{Mythril}, \cite{OsirisTool}, \cite{OyenteTool}, \cite{SecurifyTool}, \cite{SlitherTool}), and only three of them can work normally, and the remaining three (\cite{MaianTool}, \cite{OsirisTool}, \cite{OyenteTool}) cannot analyze the smart contracts of Solidity 0.4.26 and subsequent versions, because they have not been updated for a long time, thus no bugs are detected. This also reflects a real problem: when the Ethereum virtual machine (EVM) is updated, due to lack of update and maintenance motivation, the bytecode-based tools may not be able to adapt to the new version of EVM, resulting in the limited availability of these tools. However, since bytecode-based analysis tools usually use techniques such as control flow (data flow) analysis~\cite{5504796}, they usually have higher precision.


\subsection{Result analysis}
Fig~\ref{fig:exp_data} shows that \emph{Slither} detects most kinds of bugs and has a good recall and precision rate. 
The analysis tool with the highest precision is \emph{Mythril}, and \emph{Remix} has the highest recall rate. Therefore, we recommend to use \emph{Mythril}, \emph{Slither}, and \emph{Remix} for contract analysis, and the installations of these three tools are also very convenient (\emph{Slither}: through \emph{pip3}~\cite{Pip} installation, \emph{Mythril}: through free plug-in in \emph{Remix IDE}~\cite{Remix}, and \emph{Remix}: through \emph{Remix IDE}).

Our evaluation results are consistent with those in existing studies on evaluating smart contract analysis tools~\cite{durieux2019empirical, parizi2018empirical, asem2020how}. \emph{Mythril} and \emph{Slither} are generally regarded as the most effective tools. Our evaluation also has other interesting findings: \emph{i)} The crafted contracts can make these nine tools produce more false positives and false negatives (compared to the detection of non-crafted contracts). Among them, the crafted contracts that were made according to \emph{strategy} 1 make all static code scanning tools have lower recall rate. The contracts that were constructed according to \emph{strategy 2} make all the evaluated tools have lower precision rate. The crafted  contracts that were built according to \emph{strategy 3} made some evaluated tools have lower precision rate. \emph{ii)} None of these nine tools can detect the following 10 kinds of bugs: \emph{A-b-I}, \emph{D-b-P}, \emph{E-a-H}, \emph{E-a-SA}, \emph{E-a-SW}, \emph{F-a-S}, \emph{F-b-DBC}, \emph{G-a-II}, \emph{H-a-R}, \emph{H-b-N}.
\emph{iii)} Although most analysis tools currently use techniques such as control flow (or data flow) analysis, static code scanning tools are still valuable, 
because it is relatively easy to develop static code scanning tools and they can detect more kinds of bugs with better coverage. Moreover, they are usually less affected by the new versions of \emph{EVM} and \emph{Solidity}. 

\section{Related work}
\subsection{Statistics and classification of smart contract bugs}
Some studies focus on the statistics and classification of smart contract bugs. 
Delmolino et al.~\cite{delmolino2016step} summarize four common smart contract programming pitfalls by investigating students' mistakes in learning smart contract programming. Atezi et al.~\cite{atzei2017survey} summarize 11 kinds of programming traps that may lead to security bugs. They believe that one of the main reasons for the continuous proliferation of smart contract bugs is the lack of inductive documentation for smart contract bugs. Chen et al.~\cite{9072659} collect smart contracts from \emph{Stack Exchange} and Ethereum, define 20 kinds of code smell for smart contracts through manual analysis of smart contracts. 
Through interviews with smart contract developers, Zou et al.~\cite{zou2019smart} reveal that smart contract developers still face many challenges when developing contracts, such as rudimentary development tools, limited programming languages, and difficulties in dealing with performance issue. Sayeed et al.~\cite{8976179} divide the attacks on Ethereum smart contracts into four categories according to the attack principle and introduce 7 kinds of smart contract bugs, and then they provide suggestions for implementing secure smart contracts. 
Dingman et al.~\cite{8886793} first count the existing bugs of Ethereum smart contract and then classify them using the \emph{NIST} framework. They count 49 kinds of bugs and then classify 24 of them. Tikhomirov et al.~\cite{tikhomirov2018smartcheck} divide 20 kinds of smart contract bugs into security, functional, operational, and developmental, and give the severity of various bugs. 
\emph{Smartdec}~\cite{smartdec} divides the Ethereum smart contract bugs into three major categories: blockchain, language, and model. 
Their classification covers a total of 33 kinds of smart contract bugs. However, these studies only introduce partial kinds of bugs and corresponding Detection criteria.


\subsection{Smart contract datasets}
Some organizations and researchers provide datasets of smart contracts with bugs. \emph{SmartContractSecurity} provides a list of smart contract bugs, including 33 kinds of bugs and buggy smart contracts. However, \emph{SmartContractSecurity} does not classify these bugs, and some kinds of bugs also lack sample smart contracts~\cite{SWCRegistry}. \emph{crytic} provides some examples of \emph{Solidity} security issues covering 12 kinds of bugs, but 11 of them have not been updated for two years~\cite{NotSoSmartContracts}. 
Durieux et al.~\cite{durieux2019empirical} collect 47,587 Ethereum smart contracts, and then manually mark the smart contract bugs in 69 of these contracts. Based on the smart contract bug classification provided by \emph{DASP}~\cite{DASP}, they divided the smart contract bugs in 69 contracts into ten categories. But these smart contract datasets do not cover all kinds of smart contract bugs, and the number of smart contracts in them is relatively small.

\subsection{Evaluating smart contract analysis tools}
Some studies focuses on evaluating the abilities of available smart contract analysis tools. Parizi et al.~\cite{parizi2018empirical} evaluate four smart contract analysis tools using several commonly used datasets. 
Durieux et al.~\cite{durieux2019empirical} develop an execution framework \emph{SmartBugs} containing the smart contract analysis tools, and then they use \emph{SmartBugs} to evaluate 9 smart contract analysis tools. 
Ghaleb et al.~\cite{asem2020how} implement \emph{SolidFI}, a systematic method for automatically evaluating smart contract analysis tools. \emph{SolidFI} first injects bugs into the contract, then runs smart contract analysis tools to detect bugs, and finally identifies false positives and omissions generated by the tool. Our evaluation results are similar to these work. Furthermore, as a complement to existing research, we use the crafted contracts as the test cases and also evaluate the performance of \emph{Remix}. By doing so, we have more interesting and useful observations.

\section{conclusion}
By collecting all known bugs in smart contracts, we classify them based on the extension of \emph{IEEE Standard Classification for Software Anomalies} and give the detect criterion for each kind of bugs. Moreover, we construct a comprehensive dataset of vulnerable smart contracts that covers all these bugs, and obtained new observations by using this dataset to evaluate the state-of-the-art tools for analyzing smart contracts. In future work, we will study how to formally define detection criteria.

\section{Acknowledgement}
The work is is partially supported by the Natural Science Foundation of Jiangsu Province under Grant No. BK20191297, the National Natural Science Foundation of China under Grant No. 61572171, and Hong Kong RGC Project (No. 152193/19E).



\bibliographystyle{IEEEtran}
\bibliography{JiuzhouBib}

\begin{thebibliography}{10}
\providecommand{\url}[1]{#1}
\csname url@samestyle\endcsname
\providecommand{\newblock}{\relax}
\providecommand{\bibinfo}[2]{#2}
\providecommand{\BIBentrySTDinterwordspacing}{\spaceskip=0pt\relax}
\providecommand{\BIBentryALTinterwordstretchfactor}{4}
\providecommand{\BIBentryALTinterwordspacing}{\spaceskip=\fontdimen2\font plus
\BIBentryALTinterwordstretchfactor\fontdimen3\font minus
  \fontdimen4\font\relax}
\providecommand{\BIBforeignlanguage}[2]{{%
\expandafter\ifx\csname l@#1\endcsname\relax
\typeout{** WARNING: IEEEtran.bst: No hyphenation pattern has been}%
\typeout{** loaded for the language `#1'. Using the pattern for}%
\typeout{** the default language instead.}%
\else
\language=\csname l@#1\endcsname
\fi
#2}}
\providecommand{\BIBdecl}{\relax}
\BIBdecl

\bibitem{atzei2017survey}
N.~Atzei, M.~Bartoletti, and T.~Cimoli, ``A survey of attacks on ethereum smart
  contracts (sok),'' in \emph{International Conference on Principles of
  Security and Trust}.\hskip 1em plus 0.5em minus 0.4em\relax Springer, 2017,
  pp. 164--186.

\bibitem{understandTheDAO}
\BIBentryALTinterwordspacing
peckshield. (2020, May) Understanding \emph{The DAO} accident. [Online].
  Available:
  \url{https://medium.com/@peckshield/uniswap-lendf-me-hacks-root-cause-and-loss-analysis-50f3263dcc09}
\BIBentrySTDinterwordspacing

\bibitem{luu2016making}
L.~Luu, D.-H. Chu, H.~Olickel, P.~Saxena, and A.~Hobor, ``Making smart
  contracts smarter,'' in \emph{Proceedings of the 2016 ACM SIGSAC conference
  on computer and communications security}.\hskip 1em plus 0.5em minus
  0.4em\relax ACM, 2016, pp. 254--269.

\bibitem{kalra2018zeus}
S.~Kalra, S.~Goel, M.~Dhawan, and S.~Sharma, ``Zeus: Analyzing safety of smart
  contracts.'' in \emph{NDSS}, 2018.

\bibitem{torres2018osiris}
C.~F. Torres, J.~Sch{\"u}tte \emph{et~al.}, ``Osiris: Hunting for integer bugs
  in ethereum smart contracts,'' in \emph{Proceedings of the 34th Annual
  Computer Security Applications Conference}.\hskip 1em plus 0.5em minus
  0.4em\relax ACM, 2018, pp. 664--676.

\bibitem{nikolic2018finding}
I.~Nikoli{\'c}, A.~Kolluri, I.~Sergey, P.~Saxena, and A.~Hobor, ``Finding the
  greedy, prodigal, and suicidal contracts at scale,'' in \emph{Proceedings of
  the 34th Annual Computer Security Applications Conference}.\hskip 1em plus
  0.5em minus 0.4em\relax ACM, 2018, pp. 653--663.

\bibitem{chen2017under}
T.~Chen, X.~Li, X.~Luo, and X.~Zhang, ``Under-optimized smart contracts devour
  your money,'' in \emph{2017 IEEE 24th International Conference on Software
  Analysis, Evolution and Reengineering (SANER)}.\hskip 1em plus 0.5em minus
  0.4em\relax IEEE, 2017, pp. 442--446.

\bibitem{gasreducer}
T.~Chen, Z.~Li, H.~Zhou, J.~Chen, X.~Luo, X.~Li, and X.~Zhang, ``Towards saving
  money in using smart contracts,'' in \emph{In Proc. IEEE/ACM International
  Conference on Software Engineering}, 2018.

\bibitem{gaschecker}
T.~Chen, Y.~Feng, Z.~Li, H.~Zhou, X.~Luo, X.~Li, X.~Xiao, J.~Chen, and
  X.~Zhang, ``Gas{C}hecker: Scalable analysis for discovering gas-inefficient
  smart contracts,'' \emph{IEEE Transactions on Emerging Topics in Computing},
  2020.

\bibitem{jiang2018contractfuzzer}
B.~Jiang, Y.~Liu, and W.~Chan, ``Contractfuzzer: Fuzzing smart contracts for
  vulnerability detection,'' in \emph{Proceedings of the 33rd ACM/IEEE
  International Conference on Automated Software Engineering}.\hskip 1em plus
  0.5em minus 0.4em\relax ACM, 2018, pp. 259--269.

\bibitem{tikhomirov2018smartcheck}
S.~Tikhomirov, E.~Voskresenskaya, I.~Ivanitskiy, R.~Takhaviev, E.~Marchenko,
  and Y.~Alexandrov, ``Smartcheck: Static analysis of ethereum smart
  contracts,'' in \emph{2018 IEEE/ACM 1st International Workshop on Emerging
  Trends in Software Engineering for Blockchain (WETSEB)}.\hskip 1em plus 0.5em
  minus 0.4em\relax IEEE, 2018, pp. 9--16.

\bibitem{grishchenko2018ethertrust}
I.~Grishchenko, M.~Maffei, and C.~Schneidewind, ``Ethertrust: Sound static
  analysis of ethereum bytecode,'' \emph{Technische Universit{\"a}t Wien, Tech.
  Rep}, 2018.

\bibitem{liu2018reguard}
C.~Liu, H.~Liu, Z.~Cao, Z.~Chen, B.~Chen, and B.~Roscoe, ``Reguard: finding
  reentrancy bugs in smart contracts,'' in \emph{2018 IEEE/ACM 40th
  International Conference on Software Engineering: Companion
  (ICSE-Companion)}.\hskip 1em plus 0.5em minus 0.4em\relax IEEE, 2018, pp.
  65--68.

\bibitem{tokenscope}
T.~Chen, Y.~Zhang, Z.~Li, X.~Luo, T.~Wang, R.~Cao, X.~Xiao, and X.~Zhang,
  ``Tokenscope: Automatically detecting inconsistent behaviors of
  cryptocurrency tokens in ethereum,'' in \emph{Proceedings of the 2019 ACM
  SIGSAC Conference on Computer and Communications Security}, 2019, pp.
  1503--1520.

\bibitem{8728953}
J.~{Ye}, M.~{Ma}, T.~{Peng}, Y.~{Peng}, and Y.~{Xue}, ``Towards automated
  generation of bug benchmark for smart contracts,'' in \emph{2019 IEEE
  International Conference on Software Testing, Verification and Validation
  Workshops (ICSTW)}, 2019, pp. 184--187.

\bibitem{8886793}
W.~{Dingman}, A.~{Cohen}, N.~{Ferrara}, A.~{Lynch}, P.~{Jasinski}, P.~E.
  {Black}, and L.~{Deng}, ``Classification of smart contract bugs using the
  nist bugs framework,'' in \emph{2019 IEEE 17th International Conference on
  Software Engineering Research, Management and Applications (SERA)}, May 2019,
  pp. 116--123.

\bibitem{9072659}
J.~{Chen}, X.~{Xia}, D.~{Lo}, J.~{Grundy}, X.~{Luo}, and T.~{Chen}, ``Defining
  smart contract defects on ethereum,'' \emph{IEEE Transactions on Software
  Engineering}, pp. 1--1, 2020.

\bibitem{durieux2019empirical}
T.~Durieux, J.~F. Ferreira, R.~Abreu, and P.~Cruz, ``Empirical review of
  automated analysis tools on 47,587 ethereum smart contracts,'' \emph{2020
  IEEE/ACM 42th International Conference on Software Engineering}, 2020.

\bibitem{smartdec}
\BIBentryALTinterwordspacing
smartdec. (2020, Mar.) classification. [Online]. Available:
  \url{https://github.com/smartdec/classification}
\BIBentrySTDinterwordspacing

\bibitem{SWCRegistry}
\BIBentryALTinterwordspacing
SmartContractSecurity. (2020, Jan.) Smart contract weakness classification and
  test cases. [Online]. Available: \url{https://swcregistry.io/}
\BIBentrySTDinterwordspacing

\bibitem{NotSoSmartContracts}
\BIBentryALTinterwordspacing
T.~of~Bits. (2020, Jan.) Examples of solidity security issues. [Online].
  Available: \url{https://github.com/crytic/not-so-smart-contracts}
\BIBentrySTDinterwordspacing

\bibitem{GithubWiki}
\BIBentryALTinterwordspacing
Ethereum. (2020, Mar.) The ethereum wiki. [Online]. Available:
  \url{https://github.com/ethereum/wiki}
\BIBentrySTDinterwordspacing

\bibitem{EIP}
\BIBentryALTinterwordspacing
------. (2020, Mar.) Ethereum improvement proposals. [Online]. Available:
  \url{https://eips.ethereum.org}
\BIBentrySTDinterwordspacing

\bibitem{SolidityDocument}
\BIBentryALTinterwordspacing
------. (2020, Jan.) The development documents of solidity. [Online].
  Available: \url{https://solidity.readthedocs.io/en/v0.6.2/}
\BIBentrySTDinterwordspacing

\bibitem{buterin2013ethereum}
V.~Buterin \emph{et~al.}, ``Ethereum white paper,'' \emph{GitHub repository},
  pp. 22--23, 2013.

\bibitem{chen2020understanding}
T.~Chen, Z.~Li, Y.~Zhu, J.~Chen, X.~Luo, J.~C.-S. Lui, X.~Lin, and X.~Zhang,
  ``Understanding ethereum via graph analysis,'' \emph{ACM Transactions on
  Internet Technology (TOIT)}, vol.~20, no.~2, pp. 1--32, 2020.

\bibitem{chen2017ispec}
T.~Chen, X.~Li, Y.~Wang, J.~Chen, Z.~Li, X.~Luo, M.~H. Au, and X.~Zhang, ``An
  adaptive gas cost mechanism for ethereum to defend against under-priced dos
  attacks,'' in \emph{International Conference on Information Security Practice
  and Experience}, 2017.

\bibitem{isdw20101044}
I.~Group \emph{et~al.}, ``1044-2009-ieee standard classification for software
  anomalies,'' \emph{IEEE, New York}, 2010.

\bibitem{ACM}
\BIBentryALTinterwordspacing
A.~for Computing~Machinery. (2020, Mar.) Acm digitai library. [Online].
  Available: \url{https://dl.acm.org/}
\BIBentrySTDinterwordspacing

\bibitem{IEEE}
\BIBentryALTinterwordspacing
I.~of~Electrical and E.~Engineers. (2020, Mar.) Ieee digital library. [Online].
  Available: \url{https://ieeexplore.ieee.org/Xplore/home.jsp}
\BIBentrySTDinterwordspacing

\bibitem{EthereumHomePage}
\BIBentryALTinterwordspacing
Ethereum. (2020, Mar.) Ethereum github homepage. [Online]. Available:
  \url{https://github.com/ethereum}
\BIBentrySTDinterwordspacing

\bibitem{EthereumBlog}
\BIBentryALTinterwordspacing
------. (2020, Mar.) Ethereum foundation blog. [Online]. Available:
  \url{https://blog.ethereum.org/}
\BIBentrySTDinterwordspacing

\bibitem{EthereumGitter}
\BIBentryALTinterwordspacing
------. (2020, Mar.) Ethereum chatroom. [Online]. Available:
  \url{https://gitter.im/orgs/ethereum/rooms/}
\BIBentrySTDinterwordspacing

\bibitem{parizi2018empirical}
R.~M. Parizi, A.~Dehghantanha, K.-K.~R. Choo, and A.~Singh, ``Empirical
  vulnerability analysis of automated smart contracts security testing on
  blockchains,'' in \emph{Proceedings of the 28th Annual International
  Conference on Computer Science and Software Engineering}.\hskip 1em plus
  0.5em minus 0.4em\relax IBM Corp., 2018, pp. 103--113.

\bibitem{Github}
\BIBentryALTinterwordspacing
G.~Inc. (2019, Dec.) Open-source project repository. [Online]. Available:
  \url{https://github.com/}
\BIBentrySTDinterwordspacing

\bibitem{Remix}
\BIBentryALTinterwordspacing
Ethereum. (2020, Mar.) Ethereum ide and tools for the web. [Online]. Available:
  \url{http://remix.ethereum.org/}
\BIBentrySTDinterwordspacing

\bibitem{EIP150}
\BIBentryALTinterwordspacing
V.~Buterin. (2020, Mar.) Gas cost changes for io-heavy operations. [Online].
  Available: \url{https://eips.ethereum.org/EIPS/eip-150}
\BIBentrySTDinterwordspacing

\bibitem{siegel2016understanding}
D.~Siegel, ``Understanding the dao attack,'' \emph{Retrieved June}, vol.~13, p.
  2018, 2016.

\bibitem{doingblock}
\BIBentryALTinterwordspacing
Doingblock. (2020, Mar.) smart contract security. [Online]. Available:
  \url{https://github.com/doingblock/smart-contract-security}
\BIBentrySTDinterwordspacing

\bibitem{secBit}
\BIBentryALTinterwordspacing
sec\ bit. (2020, Mar.) awesome-buggy-erc20-tokens. [Online]. Available:
  \url{https://github.com/sec-bit/awesome-buggy-erc20-tokens/blob/master/ERC20_token_issue_list.md}
\BIBentrySTDinterwordspacing

\bibitem{tsankov2018securify}
P.~Tsankov, A.~Dan, D.~Drachsler-Cohen, A.~Gervais, F.~Buenzli, and M.~Vechev,
  ``Securify: Practical security analysis of smart contracts,'' in
  \emph{Proceedings of the 2018 ACM SIGSAC Conference on Computer and
  Communications Security}.\hskip 1em plus 0.5em minus 0.4em\relax ACM, 2018,
  pp. 67--82.

\bibitem{404Checklist}
\BIBentryALTinterwordspacing
knownsec. (2020, Jan.) Ethereum smart contracts security checklist from
  knownsec 404 team. [Online]. Available:
  \url{https://github.com/knownsec/Ethereum-Smart-Contracts-Security-CheckList}
\BIBentrySTDinterwordspacing

\bibitem{ethernaut}
\BIBentryALTinterwordspacing
OpenZeppelin. (2020, Jan.) Web3/solidity based wargame. [Online]. Available:
  \url{https://ethernaut.openzeppelin.com/}
\BIBentrySTDinterwordspacing

\bibitem{capture}
\BIBentryALTinterwordspacing
SMARX. (2020, Mar.) Warmup. [Online]. Available:
  \url{https://capturetheether.com/challenges/}
\BIBentrySTDinterwordspacing

\bibitem{asem2020how}
A.~Ghaleb and K.~PattabiramanD, ``How effective are smart contract analysis
  tools ? evaluating smart contract static analysis tools using bug
  injection,'' \emph{2020 ACM 29th International Symposium on Software Testing
  and Analysis (ISSTA)}, 2020.

\bibitem{MaianTool}
\BIBentryALTinterwordspacing
MAIAN-tool. (2020, Apr.) Maian: automatic tool for finding trace
  vulnerabilities in ethereum smart contracts. [Online]. Available:
  \url{https://github.com/MAIAN-tool/MAIAN}
\BIBentrySTDinterwordspacing

\bibitem{Mythril}
\BIBentryALTinterwordspacing
ConsenSys. (2020, Jan.) Security analysis tool for evm bytecode. [Online].
  Available: \url{https://github.com/ConsenSys/mythril}
\BIBentrySTDinterwordspacing

\bibitem{OsirisTool}
\BIBentryALTinterwordspacing
christoftorres. (2020, Apr.) A tool to detect integer bugs in ethereum smart
  contracts. [Online]. Available:
  \url{https://github.com/christoftorres/Osiris}
\BIBentrySTDinterwordspacing

\bibitem{OyenteTool}
\BIBentryALTinterwordspacing
melonproject. (2020, Apr.) An analysis tool for smart contracts. [Online].
  Available: \url{https://github.com/melonproject/oyente}
\BIBentrySTDinterwordspacing

\bibitem{SecurifyTool}
\BIBentryALTinterwordspacing
ChainSecurity. (2020, Apr.) Asecurity scanner for ethereum smart contracts.
  [Online]. Available: \url{https://securify.chainsecurity.com/}
\BIBentrySTDinterwordspacing

\bibitem{SlitherTool}
\BIBentryALTinterwordspacing
crytic. (2020, Apr.) Static analyzer for solidity. [Online]. Available:
  \url{https://github.com/crytic/slither}
\BIBentrySTDinterwordspacing

\bibitem{SmartCheckTool}
\BIBentryALTinterwordspacing
smartdec. (2020, Apr.) a static analysis tool that detects vulnerabilities and
  bugs in solidity programs. [Online]. Available:
  \url{https://tool.smartdec.net/}
\BIBentrySTDinterwordspacing

\bibitem{RemixIDE}
\BIBentryALTinterwordspacing
Ethereum. (2020, Jan.) Browser-only ethereum ide and runtime environment.
  [Online]. Available: \url{https://remix.ethereum.org/}
\BIBentrySTDinterwordspacing

\bibitem{SolidityCheckTool}
\BIBentryALTinterwordspacing
xf97. (2020, Apr.) Soliditycheck is a static code problem detection tool.
  [Online]. Available: \url{https://github.com/xf97/SolidityCheck}
\BIBentrySTDinterwordspacing

\bibitem{5504796}
E.~J. {Schwartz}, T.~{Avgerinos}, and D.~{Brumley}, ``All you ever wanted to
  know about dynamic taint analysis and forward symbolic execution (but might
  have been afraid to ask),'' in \emph{2010 IEEE Symposium on Security and
  Privacy}, 2010, pp. 317--331.

\bibitem{Pip}
\BIBentryALTinterwordspacing
T.~P. community. (2020, May) The pypa recommended tool for installing python
  packages. [Online]. Available: \url{https://pypi.org/project/pip/}
\BIBentrySTDinterwordspacing

\bibitem{delmolino2016step}
K.~Delmolino, M.~Arnett, A.~Kosba, A.~Miller, and E.~Shi, ``Step by step
  towards creating a safe smart contract: Lessons and insights from a
  cryptocurrency lab,'' in \emph{International Conference on Financial
  Cryptography and Data Security}.\hskip 1em plus 0.5em minus 0.4em\relax
  Springer, 2016, pp. 79--94.

\bibitem{zou2019smart}
W.~Zou, D.~Lo, P.~S. Kochhar, X.-B.~D. Le, X.~Xia, Y.~Feng, Z.~Chen, and B.~Xu,
  ``Smart contract development: Challenges and opportunities,'' \emph{IEEE
  Transactions on Software Engineering}, 2019.

\bibitem{8976179}
S.~{Sayeed}, H.~{Marco-Gisbert}, and T.~{Caira}, ``Smart contract: Attacks and
  protections,'' \emph{IEEE Access}, vol.~8, pp. 24\,416--24\,427, 2020.

\bibitem{DASP}
\BIBentryALTinterwordspacing
N.~Group. (2020, Mar.) Decentralized application security project. [Online].
  Available: \url{https://dasp.co/}
\BIBentrySTDinterwordspacing

\end{thebibliography}

\end{document}